%Paper: hep-ph/9307294
%From: dudas@qcd.ups.circe.fr (dudas)
%Date: Mon, 19 Jul 93 15:43:11 GMT

\magnification 1200
\hsize=20truecm \vsize=22truecm
\hoffset=0.1875in
\voffset=2truecm
\tolerance=1000\hfuzz=2pt

\baselineskip=16pt
\hsize=15truecm \hfuzz=5pt
\def\carre{\hbox{\rlap{$\sqcap$}$\sqcup$}}

\vglue 3cm
\centerline {\bf Composite Supersymmetric Axion-Dilaton-Dilatino System }
\centerline {\bf and The Breaking of Supersymmetry}
\centerline {E.A. Dudas}
\centerline {Laboratoire de Physique Th\'eorique et Hautes
Energies\footnote*{Laboratoire associ\'e au Centre National de la Recherche
Scientifique}}
\centerline {Universit\'
e de Paris-Sud, B\^at. 211, 91405 Orsay, France}
\vskip 1cm
PACS numbers : 11.30Pb, 12.50Lr
\vskip 1cm
\centerline { ABSTRACT }

The spontaneous breakdown of the scale, the chiral and the superconformal
symmetries for a hidden $SU(N)$ gauge group is studied in an effective
lagrangean approach. The relevant low-energy degrees of freedom are taken
to be the composite Goldstone particles associated with these three broken
symmetries. Supersymmetry is spontaneously broken in the large $N$ limit and
soft
breaking terms in the observable sector are generated, together with
nonrenormalisable Nambu-Jona-Lasinio type interactions.

\vfill
LPTHE 93/23  (June 1993)
\eject

\beginsection{$\underline {\hbox {\bf 1 - Introduction}}$}

The dilaton and the axion were introduced in  field theory for a priori two
very
different reasons.

The axion was first postulated to exist by Weinberg and Wilzcek [1] in order to
solve
the famous $U(1)$ problem in QCD and is usually seen as the Goldstone boson of
the
$U(1)$ Peccei-Quinn type symmetries [1] which allow us to dynamically set
$\theta_{QCD} = 0$ by a chiral rotation. Its mass and couplings have been
extensively studied [2] and the experimental searches impose a rather strong
constraint on the scale $\Lambda$ associated with the spontaneous breakdown of
the $U(1)$ symmetry $10^8 GeV <\Lambda< 10^{12} GeV$.

To obtain such values in a natural way (avoiding a new hierarchy problem) the
composite models of axions [3] seem to be the most simple and elegant. A new
gauge group is introduced with the corresponding coupling constant becoming
strong at the scale $\Lambda$ producing fermion condensation and the breaking
of
the Peccei-Quinn symmetry. These models need new exotic coloured fermions
whose only manifestation at low energies is through the composite axion.

The dilaton was originally introduced with the hope of understanding the
cosmological constant problem, i.e. why the cosmological constant postulated by
Einstein in the general relativity field equations is so small [4]. Even if it
fails in
this respect it has remarkable properties worth to be studied. At the classical
level it restores the dilatation invariance and is the Goldstone boson
associated
with the spontaneaous breakdown of this symmetry.

In a supersymmetric theory the two symmetries seem to be very close in the
sense that the corresponding improved currents, together with the
supersymmetry current transform into themselves under supersymmetry and form a
realsupermultiplet of
currents [5].

The corresponding charges together with  Poincar\'e,  supersymmetry and
 conformal generators give rise to the superconformal algebra [6]. Classically
the
superconformal group is a symmetry in the limit of  zero masses and is plagued
with
quantum anomalies [7].

Introducing a common scale $\Lambda$ for the dynamical breaking of the three
symmetries and remembering that the fermionic current was the supersymmetry
current, $\Lambda$ will correspond to the spontaneous breaking of supersymmetry
with the corresponding Goldstone fermion called dilatino. The easiest way to
break supersymmetry in such a way that the soft breaking terms are in the 1TeV
region is to suppose that supersymmetry is broken in a hidden sector
communicating very
weakly with the observable one [8]. This scenario appears naturally in the
heterotic superstrings models [9], in which the hidden sector couples through
gravitational interactions.

The dilaton-axion system arises naturally in any superstring inspired model in
the
$d=10$ supergravity multiplet and in the compactification process as internal
manifold degrees of freedom [10].

The purpose of this paper is to construct a low energy-effective lagrangian of
the
observable matter coupled to the axion-dilaton-dilatino
 system in a composite
model where they are dynamically generated composite particles.

Consider a $N=1$ supergravity theory containing a hidden and an observable
sector communicating only through gravitational interactions. The running of
the
hidden gauge group coupling constant will induce  gaugino  condensates
$<\lambda
\lambda >$ by nonperturbative effects at a scale $\Lambda$ [8]. The three
symmetries which mix the observable and the hidden sector, namely $U(1)_R$,
dilatation and  supersymmetry transformations will be dynamically broken.

The corresponding Goldstone particles, the axion, the dilaton and the dilatino
will be the only relevant degrees of freedom containing hidden gauge group
gauginos
at low energies. They will be composite bound states generated at the scale
$\Lambda$ and their interactions will contain $\Lambda$ as the fundamental
scale.

The other Goldstone particles associated with the breaking of the global
symmetries concerning only the hidden gauge group have negligible interactions
with the observable sector involving negative powers of the Planck mass.

The essential point is that the only remnant memory of the hidden sector is
contained in the Goldstone particles related to the spontaneous breaking of
those
global continuous symmetries which are $\underline{\hbox{common}}$ to the
observable and the hidden sector.
 It must be emphasized that this results in a different parametrization of the
 gaugino condensation comparing with  ref.[11] The gaugino
 condensates give a v.e.v. to the auxiliary component of a composite superfield
 and will be a sign for the susy breaking.

Because the scale $\Lambda$ will be of the order $10^{11}GeV$ the supergravity
multiplet will play no role in the dynamics and we will consider a flat
gravitational
background.

Even if it is not directly implied in the dynamics we must suppose an
underlying
$N=1$ supergravity, otherwise the observable and the hidden sector do not
communicate and we will have two separate $U(1)_R$ symmetries. As a
consequence in this case the breaking of the hidden sector $U(1)_R$ and
of the corresponding supersymmetry will not be transferred to the
observable part.
Section 2 introduces the supercurrent and the superconformal anomalies. A very
simple proof is presented in the appendix which shows that the Wess-Zumino
chiral field
contribution to the anomalies is chiral to all orders in the perturbation
theory.
Section 3 gives a supersymmetric definition of the three Goldstone particles,
studies the transformation properties under the various symmetries and
constructs
an effective lagrangean at low energies which reproduces the symmetry behaviour
of the high-energy theory in the spirit of ref. [11].

Section 4 deals with the analysis of  supersymmetry breaking in this
model in the large $N$ limit, the
generation of the soft-breaking terms and the radiative stabilisation of the
dilaton potential. The problem of the cosmological constant is discussed.

In the conclusion some speculations about possible dynamical effects
at low energies of the nonrenormalisable terms  induced
at the scale $\Lambda$ are given.
\beginsection{$\underline {\hbox{\bf 2- The supercurrent and the superconformal
anomalies}}$}

The currents corresponding to the three above symmetries can be put in a real
superfield ${\cal V}_m$ [5] with the following components :

$$\eqalignno{
V_n &= C_n^5 + i \theta{\chi}_n - i {\bar \theta} {\bar \chi}_n + {i \over 2}
\theta^2 [M_n + iN_n] \cr
& -{i \over 2} {\bar \theta}^2 [M_n - i N_n] - \theta \sigma^m {\bar \theta}
{\cal
V}_{m n} + i \theta^2 {\bar \theta} [{\bar \lambda}_n +{i \over 2} {\bar
\sigma}^m
\partial_m \chi_n]\cr
&- i {\bar \theta}^2 \theta [\lambda_n + {i \over 2} \sigma^m \partial_m {\bar
\chi}_n] + {1 \over 2} \theta^2 {\bar \theta}^2 [D_n + {1 \over 2} \carre
C_n^5] &(1)
\cr}$$

We define ${\cal V}_n = {\bar \sigma}_n ^{\dot \alpha \alpha} {\cal V}_{\alpha
\dot \alpha}$ where $\alpha$ and $\dot \alpha$ are two-component indices. The
divergence of the supercurrent ${\cal V}_n$ has anomalies [7] which can be
summarized in the following formula
$$
D^\alpha {\cal V}_{\alpha \dot\alpha} = {\bar D}_{\dot \alpha} {\cal A}^+
\eqno(2)
$$
where ${\cal A}$ is a chiral superfield. A simple proof of  eq.(2) will be
presented in the appendix in a very convenient form.
 Eq.(2) tells us that ${\cal V}_n$ must obey the constraint
$$
\bar D_{\dot \beta} D^2 {\cal V}_{\alpha \dot\alpha} = 4
i(\sigma^m)_{\alpha \dot
\alpha} \partial_m D^\beta {\cal V}_{\beta \dot\beta} \eqno(3)
$$
which allows  to eliminate $\lambda_n , D_n$ and the antisymmetric part of
${\cal
V}_{mn}$.

The explicit form of ${\cal V}_n$ for a supersymmetric theory containing
massless
chiral $\phi$ and gauge fields  $V$  is given by
$$
{\cal V}_{\alpha \dot \alpha} = i \phi \sigma^\mu_{\alpha \dot \alpha} \
{\buildrel
\leftrightarrow\over {\partial_\mu}} \phi^\star + {1 \over 2} D_\alpha \phi
{\bar
D}_{\dot \alpha} \phi^\star + {3 \over 2} Tr(e^V W_\alpha e^{-V} {\bar W}_{\dot
\alpha}) \eqno(4)
$$
Comparing the formulae (4) and (1) and using (3) we can make the identification
$$\eqalignno{
J^{(5)}_n &= {1 \over 2} C_n\cr
J_m  &= (\chi_n + 2 \sigma_n\bar \sigma^{n} \chi_n) &(5)\cr
\theta_{mn} &= {1 \over 6} ({\cal V}_{mn} + {\cal V}_{nm} - 2 \eta_{mn} {\cal
V}^k
_k)  }
$$
where  $J^{(5)}_m$, $J_m$ and $\theta_{mn}$ are respectively the
$U(1)_R$ current, the supersymmetry current and the energy-momentum tensor.

The transformation properties of the various fields under $U(1)_R $ are given
by
$$ \left \{ \matrix{
\theta ' &=\ &e^{-{3i \alpha \over 2}} \theta \cr
\phi '(\theta ') &=\ &e^{-i \alpha } \phi (\theta )  \cr
W'(\theta ') &=\ &e^{-{3i\alpha \over 2}} W(\theta ) \cr
} \right. \eqno(6) $$

where $\phi$ are the set of all chiral superfields and W all the gauge
chiral superfields, hidden or observable. Under the dilatations the
transformations are given by

$$ \left \{ \matrix{
x ' &=\ &e^{-\beta} x \cr
\theta ' &=\ &e^{-\beta/2 } \theta   \cr
\phi ' (x',\theta ') &=\ &e^\beta \phi(x,\theta)\cr
W'(x',\theta ') &=\ &e^{{3\over 2} \beta} W(x,\theta ) \cr
} \right. \eqno(7) $$

Classically, in a massless theory the corresponding currents are conserved.
Including the generators for Poincar\'e,  the supersymmetry,  conformal
transformations and the superconformal spinorial generator they form the
superconformal algebra [6].

In the quantum theory we will have anomalies conveniently described by a chiral
superfield containing $\theta^m_ m , \partial^m J^{(5)}_m$ and $\partial^m
I_m$,
where $I_m$ is the superconformal current. The corresponding charge is "the
square root" of the conformal transformations, just as the supersymmetry
generator is "the square root" of the translation generators.

Note the similarity between the following relations
$$\matrix{
\partial_m D^m &= &\theta^n_ n\cr
\partial_m I^m &= &\sigma^m {\bar J}_m\cr}\eqno(8)$$

where $D^m$ is the dilation current defined as the first moment of the
energy-momentum tensor. Similarly $I^m$ will be the first moment of the
supersymmetry current $J_m$
$$
\left\{ \matrix{
D^m &= &x_n \theta^{nm} \cr
I^m &= &(\sigma^n x_n) {\bar J}^m \cr
} \right. \eqno(9)
$$

Classically the right hand side in (8) is zero, but the quantum anomalies will
spoil
the symmetries.

A short proof will be presented in the appendix for computing these
anomalies which has the advantage of giving the chiral matter
contribution in a form directly related to the superpotential and which
is equivalent on-shell to the usual form [7].
Using  eq.(2) we can readily check the following useful formulae concerning the
anomalies
$$\matrix{
[D^\alpha ,{\bar D}^{\dot \alpha }] {\cal V}_{\alpha \dot \alpha} &= &D^2{\cal
A} + {\bar D}^2
{\cal A}^+ \cr
\{D^\alpha ,{\bar D}^{\dot \alpha }\} {\cal V}_{\alpha \dot \alpha} &= &-2i
\partial^m {\cal V}_m = D^2 {\cal A} - {\bar D}^2 {\cal A}^+ \cr
}\eqno(10)
$$
\beginsection{$\underline {\hbox{\bf 3- The classical definition of the
axion-dilaton-dilatino system and the }}$}

$\underline {\hbox{\bf symmetry transformation properties}}$

The definition that we will adopt is such that classically the three Goldstone
particles will couple to the divergences of the corresponding symmetry
currents.
Explicitly we define their symmetry transformation properties by the
following classical equations
$$\matrix{
\carre (s + s^*) &= -{2 \gamma \over M_p} \theta^m_ m \cr
\carre(s - s^*) &= {2i \gamma \over M_p} \partial^m J^5_m\cr
i {\bar \sigma}^m \partial_m \Psi & = {2 \gamma \over M_p} {\bar
\sigma}^m J_m = {2 \gamma \over
M_p} \partial_m {\bar I}^m\cr}\eqno(11)
$$

The three equations describe the dilaton  $s+s^*$,  the axion  $s-s^*$
and the dilatino  $\psi$, respectively.
When we  write the effective action below the gaugino condensation scale
$\Lambda$ we will determine  $\gamma$ and find quantum corrections to eqn.(11)
suppressed by exponentials of $1 \over \Lambda$.

We can rewrite them in a superfield language, introducing a chiral
supermultiplet  S.
Then the equations take the form
$$
S + S^+ = { \gamma \over M_p} {[D^\alpha , {\bar D}^{\dot \alpha}] \over
\carre} {\cal
V}_{\alpha \dot \alpha} \eqno(12)
$$
The underlying theory being  supergravity, the only available mass is the
Planck scale $M_p$ which should be put on the right-hand side to assure the
correct dimensions In fact, in the effective theory $M_p$ will be replaced by
$\Lambda$ as the only relevant dynamical scale.
The last of eqns.(11) reads
$$
\carre \Psi_\beta = {2 \gamma \over M_p} \partial_m J^m_\beta \eqno(13)
$$
  Moreover (12) gives the following constraint
$$
F_s = -{ \gamma  \over M_p} {1 \over \carre} (\partial^m M_m + i \partial^m
N_m)
\eqno(14)
$$
The constraint (3) tell us
that
$$\eqalign{
M_m &= \partial_m A\cr
N_m &=\partial_m B\cr
}\eqno (15)$$
with $A$ and $B$ scalar and pseudoscalar fields respectively.

Taking for the hidden sector only gauge and gaugino fields we find for this
part, using
${\cal V}_{\alpha \dot \alpha} = {3 \over 2} Tr(e^V W_\alpha e^{-V} {\bar
W}_\alpha)$
$$
A\ +\ iB = {3 \over 2}(\bar \lambda \bar \lambda) \eqno(16)
$$
so
$$
F_s = {3 \gamma \over 2M_p} (\bar \lambda \bar \lambda) \eqno(17)
$$
The gaugino condensation will produce a v.e.v.  $F_s \not = 0$ which have no
contribution from the observable part. Now we clearly see that the
superfield $S$ is very convenient to study the gaugino condensation,because
supposing
it as relevant low-energy degree of freedom will automatically connect the
gaugino
condensation to  susy breaking.

A useful way to visualise this definition and  the leading $N$ generation of
the
effective action at the condensation scale is to introduce $S$ as a static
field
coupled to the Ferrara-Zumino supercurrent. The lagrangean will be modified by
$$
{\cal L} \rightarrow {\cal L}-\ {\gamma \over M_p}\int d^4 \theta ( S +
S^+)
{[D^\alpha,\bar D^{\dot \alpha}]\over\carre}\  {\cal V}_{\alpha \dot \alpha}
\eqno(18)$$

At the scale $\Lambda$ due to the chiral symmetry breaking a supersymmetric
dynamics will be induced for $S$ of the type  ${\gamma^2 a^2\Lambda^2 \over
M^2_p}\int d^4 \theta S^+S$ , the only leading term in the $1 \over N$
expansion if  $\gamma \sim{1 \over N}$. We make a one-loop computation for the
dynamics
with the chiral symmetry breaking effect taken into account by massive
gauginos.
One obtains  $a^2 = {N^2 {m_\lambda}^2 \over
2\pi^2\Lambda^2}\ell n {\Lambda^2 \over {m_{\lambda}}^2}$, where
$m_\lambda$  is a typical  hidden sector gaugino mass. Taking into
account all the planar diagrams will add a multiplicative constant to the
above expresion which is not essential in  our analyses. Choosing
$\gamma$ as to normalize the induced kinetic
term we remark that  $\Lambda$ is the only relevant scale in the problem and
$M_p$
disappears completely from the lagrangean. Writing the field equations for $S$
we find
$$
D^2 S= {4\over a\Lambda}{ D^2[D^\alpha,\bar D^{\dot \alpha}]
                            \over \carre}
{\cal V}_{ \alpha \dot \alpha}= {64 \over a\Lambda}{\cal A}^+
$$
which is equivalent to (12) in components if we replace  ${M_p \over \gamma}$
by
$a\Lambda \over 4$.

The transformation properties of the superfields  $S$  under the $U(1)_R$ and
the
dilatation are defined by the right-hand side of (12) with the above-mentioned
replacement. Being a dynamical scale, $\Lambda$
transforms as a dimension one field under dilatations. Then  we obtain
the following results :   under $U(1)_R$
$$
\qquad S\ +\ S^+ = \hbox{inv.}, \eqno(19)
$$
and under dilatations
$$
\qquad S\ +\ S^+\ \rightarrow\ e^\beta (S\ +\ S^+). \eqno(20)
$$

 The first transformation allows us to define
$$
S\ \rightarrow\ S\ +\ ia  \Lambda \alpha \eqno(21)
$$
under $U(1)_R$ with an arbitrary real coefficient $a$ , and the second one
$$
S\ \rightarrow \ e^\beta S \eqno(22)
$$
Then denoting by  $d$ and $a$ the real dilaton and the axion, and using the
variables $y$ and $\theta$, where $y^m = x^m + i\theta\sigma^m {\bar
\theta}$,  we
can write the superfield $S$ as
$$
S(y,\theta) = i a + e^d + \theta\psi + \theta^2 F \eqno(23)
$$
In the following using the symmetries of  and the anomaly structure of
the theory above the
condensation scale $\Lambda$ we will find the effective lagrangian for the
observable matter plus the Goldstone system.

In order to have the correct coupling to matter we must have an interaction of
type
$$
{1 \over \Lambda} \int d^4 \theta  (S+S^+) {[D^\alpha,\bar D^{\dot \alpha}]
\over \carre} {\cal
V}_{\alpha \dot \alpha} = {2i \over \Lambda} \int d^4 \theta (S-S^+) {1 \over
\carre}
\partial^m {\cal V}_m \eqno(27)
$$
In (27) only the observable part is retained in the current  ${\cal V}_m$,
fact that which will be tacitly assumed in the following. Then because
all the induced lagrangian is generated from this primary interaction it
will be a function of $S$ and $\partial^m{\cal V}_m$. The first terms in
a ${1 \over\Lambda}$ expansion are of the form

$$\eqalignno{
{\cal L}_{{eff, inv}} &=
   \int d^4 \theta \left\{ S^+ S+ {i \over a\Lambda} (S-S^+) {1 \over
\carre} \partial^m {\cal V}_m +{d \over \Lambda^2} (\partial^m {\cal
V}_m)^2\right \} \cr
&+ \left (e \Lambda^3 \int d^2 \theta e^{-{3S \over a\Lambda}} + h.c.\right )
&(28)\cr }$$
The canonical kinetic term for S is fixed by the $U(1)_{R}$ symmetry and
is different from that of the  superstring-inspired dilaton. It will be
the same as those corresponding to the Standard Model dilaton [4] and
axion [1].
${\cal L}_{eff}$ must be globally supersymmetric in order to obtain a
spontaneous
breaking of SUSY which gives only soft breaking terms [15] and no quadratic
divergences. It is natural to have the condensation scale $\Lambda$ as the
fundamental mass parameter in ${\cal L}_{eff}$ because the effective
interaction
between Goldstone particles and observable matter is generated at that scale.
The third term
is the lowest nonrenormalisable generated for the observable part and we will
see
that it can give dynamical effects at low energies. The last term is the most
general
superpotential for the $S$ field invariant under $U(1)_R$ and dilatations and
$a$ is
the same coefficient appearing in the transformation of the axion (21), which
is
arbitrary for the moment and will be fixed by asking the compensation of
the $U(1)_R$ anomaly .

%We can understand the generation of the effective action in terms of the
%%linear
%superfield $L$ introduced in the eq.(12). It is readily checked that at the
%condensation scale $\Lambda$ we generate a dynamics $\sim - {1 \Lambda^2 \over
%2M_p^2} \int. d^4 \theta L^2$ which takes care of the first three terms in
%%(23) when
%eliminating $L$. The computation of all generated terms in $L$ is very
%%difficult
%however, so it is preferable to use symmetry arguments in determining ${\cal
%L}_{eff}$.

To reproduce the hidden sector anomalies at low energy we must introduce an
anomalous term of the form [11]
$$
{\cal L}_{an} = {1 \over 3} \int d^2 \theta {\cal A} (\ell n {{\cal A} \over
\mu^3} - 1) +
h.c. \eqno(29)
$$
In the above formula (29), $\mu$ does not transform under dilatations in
order to correctly reproduce the anomalies.

Using (10) and the defining eq. (12) we find classically
$$
S + S^+ = {1 \over M_p} {1 \over \carre} (D^2 {\cal A} + {\bar D}^2 {\cal A}^+)
\eqno(30)
$$
and as a consequence
$$
{\cal A} = {M_p \over 16} (\bar D^2 S^+) \eqno(31)
$$
Introducing this last eqn. in (29) we will find in fact $\delta {\cal L}_{an} =
0$ under both $U(1)_R$
and dilatations.
%It is normally because is a classical relation which do not take into
%account the quantuum induced lagrangian for  $S$, for example the last term in
%%(  ).
%Because the anormalous term will be of higher order we can use the $S$ field
%equations from ( ) to compute ${\cal A}$ and substitute it in (  ). We obtain
%%the
%following $S$ superfield equation
%$$\eqalign{
%D^2 (S - {c \over 2\Lambda}\ { [D^\alpha,\bar D^{\dot \alpha}] \over \caOAOrre
%%} {\cal
%V}_{\alpha \dot \alpha}) &= -{8e \over  bo}\ \Lambda^2.e^{{2S^+ \over bp
%\Lambda}}\cr
%\bar D^2 (S^+ - {c \over 2\Lambda}\ {[D^\alpha,\bar D^{\dot \alpha}]
%\over \carre} {\cal V}_{\alpha \dot \alpha}) &= -{8e \over  bo}\ \LamOOAbda^2
%%.e^{{2S^+
%\over bp \Lambda}}\cr } \eqno(32)
%$$
%Combining with (  ) we find
%$$
%{8c \over \Lambda} {\cal A} = \bar D^2 S^+ + {8e \over bo} \Lambda^2 . e^{-{2S
%\over bo \Lambda}} \eqno(33)
%$$
To write the effective lagrangean for $S$ we need a relation between
$S$ and ${\cal A}$ which is equivalent to an off-shell definition of the
Goldstone particles contained in $S$. Lacking any natural candidate, the
best thing to do is to find the most general one compatible with the
symmetries.
Then we are lead to the following equation
$$
{\cal A} = b \Lambda \bar D^2 S^+ + c \Lambda^3 e^{-{3S\over a\Lambda}}
\eqno(32)
$$
We will see that taking $b=0$ gives two supersymmetric minima
, so $b$ is a
crucial parameter related to the dynamical nature of the scale $\Lambda$.
Indeed regarding $\Lambda$ as a fixed scale forces us to consider
only the second term in the right-hand side of eq. (32) and  supersymmetry is
not broken.  The complete induced lagrandean is defined by the eqs. (28), (29)
and (32).
 At the one-loop level, the interaction $S$ field-observable matter will
produce an effective term equivalent to
replacing the anomaly equation (10a) into the lagrangean (28) which gives
$$
{\cal L}_{one-loop} = -{1 \over 2a\Lambda} \int d^4\theta (S-S^+)({D^2
\over\carre}
{\cal A} - {\bar D \over \carre} {\cal A}^+) = {2 \over a\Lambda} (\int d^2
\theta
S{\cal A} + \int d^2 \bar \theta S^+ {\cal A}^+) \eqno(34)
$$
We can interpret this term as a generation of a superpotential taking into
account
the form of the anomaly given in the appendix. To obtain that we need a
nontrivial v.e.v.
for the dilaton. Now we see that the transformation of $S$ under $U(1)_R$
was chosen to compensate the anomaly.
The characteristic feature of this composite model is that in addition to the
usual
interaction with the gauge fields present in the superstring inspired
supergravity
models it generates an interaction with the superpotential. In components we
will
obtain a usual coupling of the axion-dilaton system to the matter [1-4].

We will be interested in the large $N$ limit of the theory in order to make
quantitative predictions about the supersymmetry breaking. To have a nontrivial
$1 \over N$ development we must be able to factorize the $N$ dependence
in the lagrangean by performing field rescalings. Making $W^\alpha
\rightarrow N^{1 \over 2} W^\alpha$ and $S \rightarrow N^{1 \over 2} S$ we
fulfill this requirement if the quantities  ${a \over N^{1 \over 2}}, {b
\over N},{c \over N^2}, {e \over N}$  and $dN$ are held fixed in the
large $N$ limit. To derive the large $N$ behaviour of $b$ and $c$ we
used the form of the anomaly ${\cal A}$ given in (A.16).

Interpreting $\Lambda$ as a dynamical scale, all the terms but ${\cal L}_{an}$
are
invariant under the dilatations such that the dilaton has really the role of
restoring
the symmetry at the classical level.
It should be emphasized that we are dealing with two different scale
symmetries,
a high energy one which was described in this paragraph and a low energy
one when $\Lambda$ is kept fixed. In the latter case redefining the
scale transformation of $S$ we can interpret the term in eq.(34) as being the
Wess-
Zumino term for the three anomalous symmetries.

\beginsection{$\underline{\hbox{\bf 4 - The dynamical breaking of
supersymmetry}}$}

Usually in a globally supersymmetric theory the gaugino condensation do not
breaks supersymmetry\footnote{*}{See, for example, G. Veneziano and S.
Yankielowicz in ref. [11].} and we need a v.e.v. for $\theta^\mu_\mu$ to do
that,
according to the relation
$$
< 0 | {1 \over 8} Tr \biggl \{(\bar \sigma^\mu J_\mu)^{\dot \alpha} , \bar
Q_{\dot \alpha}J\biggr \} | 0 > = < \theta^m_m > = {\beta(g) \over 2g}
<F^{a}_{mn}
F^{amn} > \eqno(35)
$$
which vanishes in globally super Yang-Mills theory. In our case however the
dynamical bound states  S  may change the situation. Using the formula
$$
< \theta^m_mJ> = < \int d^2\theta {\cal A} + \int d^2 \bar \theta {\cal A}^+>
\eqno(36)
$$
and using the identification of the chiral anomaly ${\cal A}$ in eq.(32) we
find
$$\eqalignno{
&{1 \over 8} < 0 | Tr \biggl \{ \bar \sigma_\mu J^\mu , \bar Q \biggr \} | 0 >
=
{c\Lambda^3} < \int d^2 \theta e^{-{3s \over a \Lambda}} + \int d^2
\bar \theta e^{-{3s^\star \over a \Lambda}} > =\cr
&= - {3c \over a} \Lambda^2 <e^{{-3s \over a \Lambda}} F_s + e^{{-3s^\star
\over
a\Lambda}} F^\star _s > &(37)\cr}
$$
In order to break susy we need $<e^{-{3s \over a\Lambda}} F_s > \not= 0$ at the
minimum of the effective action, a  condition sufficient for generating
soft-breaking terms in the observable sector. Writing only the terms in
(28-29)  relevant  for the  s  effective potential, we obtain

$$\eqalignno {
&{\cal L}_s = F^\star_s\ F_s - {3e \over a}\ \Lambda^2J\ \biggl ( e^{-{3s \over
a\Lambda }}\ F_s\ +\ e^{-{3s^\star \over a\Lambda }} F^\star _s \biggr ) - \cr
&- {c \Lambda^2 \over a} \biggl \{ e^{-{3s \over a \Lambda }} F_s \ell
n  \biggl ( {b \Lambda \over \mu^3} F^\star_s\ + c\ {\Lambda^3 \over
\mu^3}\ e^{-{3s \over a \Lambda }} \biggr )\ +\  e^{-{3s^\star \over a \Lambda
}}\ F^\star_s \ell n \biggl (  {b \Lambda \over \mu^3} F_s + c
\ {\Lambda^3 \over \mu^3}\ e^{-{3s^\star \over a \Lambda }} \biggr ) \biggr \}
\cr
&&(38) \cr
} $$
It is readily seen that $c$ and $e$ are redundant parameters that can be
absorbed in
redefinitions of $s$ and $\mu$ such that $c =1$ and $e =0$.
As it must be, $F_s$ is a constraint and is defined implicitly through the
following
equation

$$\eqalignno{
\biggl (1 -{b \over a} &\ {e^{-{3s^\star \over a\Lambda}}
\over{b F_s \over \Lambda^2}+  e^{-{3s^\star \over a\Lambda }}} \biggr )
F^\star_s =\cr
&= \ {1\over a} \Lambda^2 \ e^{-{3s \over a \Lambda}} \ell n \biggl ({b\Lambda
\over\mu^3} F^\star_s \ + { \Lambda^3 \over \mu^3} e^{-{3s \over a
\Lambda}} \biggr ) &(39)\cr}
 $$
Using it in (38) we obtain the  $s$ potential

$$
V = |F_s|^2 \biggl\{- {b \over a} \biggl ( {e^{-{3s^\star \over a
\Lambda}} \over {bF_s \over \Lambda^2}+  e^{-{3s^\star \over a
\Lambda}} } + {e^{-{3s \over a \Lambda}} \over{b F^\star_s \over \Lambda^2}+
e^{-{3s \over a \Lambda}} } \biggr ) + 1\biggr \} \eqno (40)
$$
%Of course an analytical study of the potential is impossible due to the
%%implicit
%equation (  ). Moreover we have three free parameters $b_o$ , $c$ and $e$. We
%%will
%content ourselves to determine them tell that we obtain a zero cosmological
%constant and a broken susy.

%First of all we remark from (  ) and (  ) that $s \rightarrow \infty$ implies
%%$F_s
%\rightarrow 0$ and $V = 0$, a rather expected result.
%Minimizing $V$ in (  ) using (  ) we obtain an expression $< V > = F_1(b_o ,
%%c, e)$.
%Putting $< V > = 0$ we obtain a relation between the parameters. Another
%%relation
%will be obtained remarking that at the minimum
%$$
%< F_s > = {8 e \Lambda^2 \over b_o} (-1 + {1 \over 6 b_o c}) e^{-{2<s^\star>
%%\over
%b_o\Lambda}} \eqno(41)
%$$
%is compatible with $< V > = 0$ and fixes $< s >$ by (  )  to be
%$$
%{2s \over b_o \Lambda} = 9 b_o c + \ell n {e \Lambda^3 \over 6 b_o^2 c^2
%%\mu^3} -
%1 \eqno(42)
%$$
%which should be compatible with the minimization of $V$. This gives a second
%equation $F_z(b_o , c , e) = 0$. So in principle we can determine two
%%coefficients
%the out of $b_o , c$ and $e$ demanding a zero cosmologiocal constant and a
%%broken
%susy.
To find the minimum we minimize $V$ with respect to $s$

$$
F_s e^{-3s \over a\Lambda} \biggl\{\ell n \biggl ( {b\Lambda F^\star_s
\over\mu^3}+{\Lambda^3 \over \mu^3} e^{-3s \over a\Lambda} \biggr ) +
{\Lambda^3 \over \mu^3} {e^{-3s \over a\Lambda} \over {{\Lambda^3 \over\mu^3}
e^{-3s \over a\Lambda} +{ b\Lambda \over \mu^3} F^\star_s}} \biggr \} =0
\eqno(41)
$$
and combine it with (39).
Searching for real solutions, we find the following results:

$$\eqalign{
i)  F_s& =0 ,  <V> =0\cr
ii-iii){ 3s \over a \Lambda}& =\ell n {\Lambda^3 ( bA + 1) \over  \mu^3}
     +{1\over (bA + 1)}\cr
    F_s \over \Lambda^2&=  A e^{ -3s \over a \Lambda}\cr
    <V>&= -{\Lambda^4 A^2} e^{-6s \over a \Lambda}
   \biggl [ {2b \over {a(bA + 1)}} -1 \biggr ]\cr } \eqno(42)
$$

where  $ A = {1 \over 2ab} \biggl [ b-a \pm {((b-a)^2 - 4ab)}^{1 \over 2}
\biggr ]$.
The case $i)$ contains the runaway vacuum obtained sending $s$ to infinity
and another supersymmetric solution defined by   $ {3s \over a \Lambda} =
\ell n {\Lambda^3 \over \mu^3}$. These are the only extrema in the case  $b
=0$.
{}From (42) we remark that the only solution for a vanishing cosmological
constant  $<V> =0$ is  $b=0$. In this case  $F_s=0$ and supersymmetry is not
broken.

To obtain a definite result for the real minimum  we will take the large $N$
limit of these equations.
We get $A = {1 \over a}$ and $A = {1 \over b}$. The first value
of $A$ is the  minimum in the large $N$ limit and it breaks supersymmetry
generating at the same time a large cosmological constant  $<V> =
-{\Lambda^4 \over a^2} e^{ -6s \over a \Lambda}$.

The soft-breaking terms are generated from the eqs.(28) and (29). Because the
 $\theta^2$ and $\bar \theta^2$ components of the eqs (10) are identities, both
of
them will have the same form
$$
{1 \over a\Lambda} F_s{A_1 \lambda \lambda + A_2 w} + h.c.\eqno(43)
$$
with $w = W$, and $A_1$ , $A_2$ some numerical coefficients directly
calculable.
 Together with the soft scalar masses which will be generated by
radiative corrections being no more protected by susy they constitute the most
general susy breaking terms which do not produce quadratic divergences [15]. As
in the usual gauging condensation scenarios [8] to obtain interesting values
for
the soft terms we need
$$
{F^\star_s \over \Lambda}  \sim {\mu^3 \over \Lambda^2} \sim 1 TeV  \eqno(44)
$$
which for $ \Lambda = 10^{11} GeV$ requires $ \mu \sim 10^8 GeV$. Then we can
check that the susy breaking solution gives $s > 0$ so the generated
superpotential
has the required sign and we have a real value for the v.e.v. of the
dilaton $d$ defined in the equation (23).

\beginsection{$\underline{\hbox {\bf Conclusions }}$}

   The composite axion-dilaton-dilatino system is used to parametrize
the low-energy degrees of freedom below the gaugino condensation scale.
We couple the corresponding superfield as a static field to the divergence of
the supercurrent as in  eq. (18). At the condensation scale  $\Lambda$  it
acquires a dynamics due
to the breaking of the chiral symmetry and becomes a propagating degree of
freedom, corresponding to the formation of bound states. Using symmetry
 considerations we have constructed an effective lagrangean  which
admits a supersymmetry breaking minimum in the large $N$ limit and
generates  soft-breaking terms for the observable sector. No fine-tuning
is possible in order to have a vanishing cosmological constant and broken
supersymmetry.

We can speculate about the possible dynamical effects induced by the third
term in the effective lagrangean (28) of the type  ${1 \over \Lambda^2}
\int d^4 \theta{(\partial_m {\cal V}_m)}^2$ . Working out the components we
find the local terms    $C^5_m C^{5m}$   and    $(A^2 + B^2)$   using the
notations
from  eqs. (1) and (15). Considering only the fermion terms we obtain
Nambu-Jona-Lasinio interactions    $(\lambda \lambda) (\bar\lambda
\bar\lambda)$
  and    ${(\psi \sigma_m \bar\psi)}^2$ ,  where   $\psi$   is some observable
fermion matter field. The second term can induce the electroweak symmetry
breaking and could explain the hierarchy between the susy breaking scale and
the electroweak scale by a top-antitop condensation mechanism [17]. The
coupling in front of this term is of the order  $G \sim {1 \over \Lambda^2}$
 and is unable to produce the condensation for values   $G <G_c \sim
{1\over \Sigma^2}$   where  $\Sigma$  is a typical value for the soft-breaking
terms. It will start to run to lower energies like an asymptotically free
coupling [18] and will reach the critical value at a much lower energy.
The explanation of the hierarchy is the same as the one between the QCD scale
and the unification scale.
   The first operator   $(\lambda\lambda) (\bar\lambda\bar\lambda)$  is
analogous
to the gravity induced interaction in supergravity and could
produce  gaugino condensation in  supersymmetric QCD.

\beginsection{$\underline{\hbox {\bf Aknowledgments}}$}

   I would like to thank P. Bin\'etruy for many enlightening discussions and
constant help. I enjoyed useful conversations with F. Pillon and C. Savoy.
\vfill\eject
\centerline{$\underline{\hbox{APPENDIX}}$}
\vskip 1cm
Consider a renormalisable theory containing arbitrary chiral multiplets
$\phi_i$
with a superpotential $W(\phi_i)$ interracting with the gauge fields of an
arbitrary
gauge group. The Lagrangian is
$$
{\cal L}=  \int d^4 \theta \phi^+ e^V \phi + \int d^2 \theta [{1 \over 4gz}
Tr(W^\alpha
W_\alpha) + W(\phi ,h_i,m_i) + h.c.\eqno (A.1)
$$
We will compute the vacuum energy density ${\cal E}_o$ which is a
renormalisation
invariant quantity depending on an arbitrary mass scale $\mu$ and the other
parameters of the lagrangean, coupling $h i$ and mass parameters $mi$.

By dimensional analysis we can write
$$
{\cal E}_o = \mu^4 f(g , h_i , {m_i \over \mu}) \eqno(A.2)
$$
where  $g$ is the gauge coupling and $f$ a dimensionless function. Being a
physical
observable, ${\cal E}_o$ obeys the renormalisation group equation
$$
[\mu {\partial \over \partial\mu} + m_i \gamma_i {\partial \over \partial m_i}
+
\beta_g {\partial \over \partial g} + \beta_i {\partial \over \partial h_i} ]
{\cal E} =
0 \eqno(A.3)
$$
Combining (A.2) and (A.3) we find the equation
$$
{\cal E} = -{1 \over 4V_4} [\beta_g {\partial \over \partial g} + m_i (1 +
\gamma_i)
{\partial \over \partial m_i} + \beta_i {\partial \over \partial h_i} ] \Gamma
\eqno(
A.4)
$$
where in the right-hand side of (A.4) we replaced ${\cal E}$ by the expression
${1
\over V_4} \Gamma$, $\Gamma$ being the Legendre transform of the connected
vacuum to vacuum generating functional and $V_4$ is the space-time volume. In
principle the Legendre transform must be performed with respect to any scalar
relevant degrees of fredom which could have vacuum expectation values, and in
(A.4)
$\Gamma$ is computed at the saddle point value. Using the definitions
$$\eqalignno{
\int {\cal D} \phi e^{i[S(\phi) + J F(\phi)]} &= e^{i W(J)}&(A.4)\cr
\Gamma(F_c) &= W(J) - J F(\phi_c)&(A.5)\cr
}$$
where $F_c = F(\phi_c) = {\delta W(J) \over \delta J}$ and
$$
J = -{\delta \Gamma (F_c) \over \delta F_c}\eqno(A.6)
$$
$\phi$ is the set of all quantum fields and $F(\phi)$ the scalar relevant
degrees of
freedom constructed out of the fundamental fields $\phi$.

At the saddle point ${\delta \Gamma \over \delta F_c} = J = 0$ so using the
defnition (A.5) we find
$$
{1 \over V_4} {\partial \Gamma (F_c) \over \partial a_i} = {1 \over V_4}
{\partial
W(J=0) \over \partial a_i} = {\partial \over \partial a_i} <{\cal L} > (A.7)
$$
where $a_i$ is any parameter of the lagrangean, namely $g, h_i$ and $m_i$ and
we
used (A.4).

The last step is to use the renormalized Schwinger principle which allow us to
act
with the derivatives inside the brackets using appropriate renormalized
expressions for the singular product of the composite operators. For example in
the $MS$ scheme using the dimensional reduction as regularisation [12]
$$
{1 \over V_4} {\partial \over \partial g} W(J=0) = -{1 \over 2g^3} \int d^2
\theta Tr
\{ N(W^\alpha W_\alpha )\} \eqno(A.8)
$$
when  $N$ denotes the renormalized product of operators in the sense of
Zimmermann [13].

After a trivial rescaling of the gauge field the final result is obtained (we
will omit
the symbol $N$ in the following, understanding that all the composite operators
are renormalized)
$$
{\cal E}_o = {1 \over 4} \int d^2 \theta < - {\beta(g) \over 2g} Tr(W^\alpha
W_\alpha) + m_i(1 + \gamma_i) {\partial W \over \partial m_i} + \beta_i
{\partial W
\over \partial h_i} > + h.c. \eqno(A.8)
$$

The result is that for the most general renormalisable lagrangean we can
express
${\cal E}_o$ as an integral over a holomorphic function. We will introduce the
chiral
field of anomalies ${\cal A}$
$$
{\cal A} = -{\beta(g) \over 2g} Tr(W^\alpha W_\alpha) + m_i \gamma_i {\partial
W
\over \partial m_i} + \beta_i {\partial W \over \partial m} \eqno(A.9)
$$
The anomalous transformations of the lagrangian under a $U(1)$ transformation
is
given by
$$
\delta_{U(1)_R} {\cal L} = i\alpha (\int d^2 \theta {\cal A} - \int d^2 \theta
{\cal
A}^+) \eqno(A.10)
$$
and under dilatations
$$
\delta {\cal L} = \beta(\int d^2 \theta {\cal A} + \int d^2 \bar \theta {\cal
A}^+)
\eqno(A.11)
$$

To check the equivalence with the usual definition of the chiral anomaly {\cal
A} [7]
we take an arbitrary gauge model with chiral superfields $\phi_i$ and the most
general renormalisable lagrangian
$$
{\cal L} = \int d^4 \theta [\phi^+_i (e^V)_{ij} \phi_i ] + \biggl(\int d^2
\theta [{1
\over 4g^2} Tr(W^\alpha W_\alpha ) + {1 \over 2} m_{ij} \phi^{i} \phi^j + {1
\over 3}
g_{ijk} \phi^{i} \phi^j \phi^k] + h.c.\biggr )
$$
Define the wave function renormalisation $Z^{i}_j$ and the renormalisation
group
$\beta_{ijk}$ , $X_{ij}$ and $\gamma_{ij}$ functions by
$$\eqalign{
\phi^{i} &= Z^{i}_j \phi^j_o\cr
X_i^\ell &= \mu {\partial \over \partial\mu} Z_i^j \cr
m_i^\ell \gamma_{\ell j} &= \mu {\partial \over \partial \mu} m_{ij}\cr
\beta_{ijk} &= \mu {\partial \over \partial \mu} g_{ijk}\cr
}\eqno(A.12)$$
The only renormalisation of $m_{ijk}$ comes from the wave-function
renormalisation due to the nonrenormalisation theorem [14]. Then we obtain in
the
lowest order of the perturbation theory
$$\eqalign{
m_i^\ell \gamma_{\ell j} &= X_i^\ell m_{\ell j} + X_j^\ell m_{\ell i}\cr
\beta_{ijk} &= X_i^\ell g_{\ell j k} + X_j^\ell  g_{i \ell k} + X_k^\ell g_{i j
\ell}\cr
}\eqno(A.13)$$
Introducing the equations (A.13 ) in the formulae for the chiral multiplet of
anomalies
$$
{\cal A} = - {1 \over 2}\ {\beta(g) \over g}\ Tr(W^\alpha W_\alpha)\ +\ {1
\over 2}
m_i^\ell \gamma_{\ell j} \phi^{i} \phi^j\ +\ {1 \over 3}\ \beta_{i j k} \phi^i
\phi^j
\phi^k $$
and using the symmetry of $m_{ij}$ and $g_{ijk}$ we obtain
$$
{\cal A} = -\ {1 \over 2}\ {\beta(g) \over g}\ Tr(W^\alpha W_\alpha)\ +
\ X^\ell_i(m_{\ell j} \phi^i \phi^j \ +\ g_{\ell i j} \phi^i \phi^j \phi^k)
\eqno(A.14)
$$

We will rewrite the chiral superfield contributions using the classical field
equations
$$
\bar D^2 [ \phi^+_k (e^V)^k_i ]\  -\ 4(m_(ij) \phi^j\ +\ g_{ijk} \phi^j \phi^k)
= 0
\eqno(A.15)
$$
with the result
$$
{\cal A} = -  {1 \over 2}\ {\beta (g) \over g}\ Tr(W^\alpha W_\alpha) + {1
\over 4}
\bar D^2 [ (\phi^+ e^V)_\ell X^\ell_i \phi^i ] \eqno(A.16)
$$
in agreement with previous results [7].

Using the formula
$$
{\cal E}_o = {1 \over 4} < \theta^m_m > \eqno(A.17)
$$
we see that $\theta^m_m$ is the real part of the highest component of the
chiral
superfield ${\cal A}$. It is easily checked that the chiral anomaly is the
corresponding imaginary part and that the $\theta$ component correspond to the
anomaly in the divergence of the superconformal current $\partial_m I^m$.

\beginsection{$\underline{\hbox{\bf References}}$}

\item{[1]} A. Peccei and H.R. Quinn, Phys. Rev. Lett. 38, 1440 (1977)  ;
\item{   }  Phys. Rev. D16, 1791 (1977) .
\item{   } S. Weinberg, Phys. Rev. Lett. 40, 223 (1978) .
\item{   } F. Wilczek, Phys. Rev. Lett. 40, 279 (1978) .
\item{[2]} H. Georgi, D.B. Kaplan and L. Randall, Phys. Lett. B169, 73 (1986) .
\item{   } P. Sikivie, Phys. Rev. Lett. 51, 1415 (1983)  ;
\item{   } Phys. Rev. Lett. 52, 695 (1984) .
\item{   } W.A. Bardeen and S-H. H. Tye, Phys. Lett. B74, 229 (1978) .
\item{[3]} J.E. Kim, Phys. Rev. D31, 1733 (1985) ;
\item{   } D.B. Kaplan, Nucl. Phys. B260, 215 (1985) .
\item{[4]} R.D. Peccei, J. Sola and C. Wetterich, Phys. Lett. B195, 183 (1987)
 ;
\item{   } W. Buchmuller and N. Dragon, Phys. Lett. B195, 417 (1987)  ;
\item{   } Nucl. Phys. B321, 207 (1989) .
\item{   } S. Weinberg, Rev. Mod. Phys. 61, 1 (1989) .
\item{[5]} S. Ferrara and B. Zumino, Nucl. Phys. B87, 207 (1975) .
\item{[6]} S. Ferrara, Nucl. Phys. B77, 73 (1974) ;
\item{   } M. Sohnius, Phys. Rep. 128, 39 (1985) .
\item{[7]} T.E. Clark, O. Piguet and K. Sibold, Nucl. Phys. B143, 445 (1978)  ;
\item{   } O. Piguet and K. Sibold, Nucl. Phys. B196, 428 (1982)  ;
\item{   } Nucl. Phys. B196, 447 (1982) ;
\item{   } P. West, Introduction to Supersymmetry and Supergravity,
World Scientific Publishing Co., Singapore, 1986.
\item{[8]} H.P. Nilles, Phys. Lett. B112, 455 (1982)  ; Nucl. Phys.
B217, 336 (1983)
\item{   } R. Barbieri, S. Ferrara and C.A. Savoy, Phys. Lett. B119, 343 (1982)
 ;
\item{   } J.P. Deredinger, L.E. Ibanez and H.P. Nilles, Phys. Lett.
B155, 65 (1985)  ;
\item{   } M. Dine, R. Rohm, N. Seiberg and E. Witten, Phys. Lett. B156, 55
(1985)
\item{[9]} D.J. Gross, J. Harvey, E. Martinec and R. Rohm, Nucl. Phys.
B256, 253 (1985)
 ;
\item{   } P. Candelas, G. Horowitz, A. Strominger and E. Witten, Nucl.
Phys. B258, 46 (1985) .
For a textbook, see M.B. Green, J.H. Schwarz and E. Witten, Superstring Theory,
 Cambridge Univ. Press, Cambridge, 1987).
\item{[10]} E. Witten, Phys. Lett. B149, 351 (1984) .
\item{[11]} G. Veneziano and S. Yankielowicz, Phys. Lett. B113, 231 (1982)  ;
\item{   } P. Bin\'etruy and M.K. Gaillard, Phys. Lett. B253 ,119 (1991)  ;
\item{   } M. Cvetic, A. Font, L. Ibanez, D. Lust and F. Quevedo, Nucl.
Phys. B361, 194 (1991) .
\item{[12]} W. Siegel, Phys. Lett. B84, 193 (1979) .
\item{   } S.J. Gates, Jr., M.T. Grisaru, M. Rocek and W. Siegel, Superspace or
One
Thousand and One lessons in supersymmetry,  Frontiers in Physics (1983).
\item{[13]} W. Zimmermann, Ann. Phys.(NY) 77, 536 (1973) .
\item{[14]} M.T. Grisaru, W. Siegel and M. Rocek, Nucl. Phys. B159, 429 (1979)
{}.
\item{[15]} L. Girardello and M.T. Grisaru, Nucl. Phys. B194, 65 (1982) .
\item{[16]} S. Ferrara, J. Wess and B. Zumino, Phys. Lett. B51, 239 (1974)  ;
\item{   } W. Siegel, Phys. Lett. B85, 333 (1979)
\item{[17]} T. Clark, S. Love and W. Bardeen, Phys. Lett. B237, 235 (1990) ;
\item{    } M. Carena, T. Clark, C. Wagner, W. Bardeen and K. Sasaki,
    Nucl. Phys. B369, 33 (1992)
\item{[18]} P. Bin\'etruy, E.A. Dudas and F. Pillon, Orsay preprint
LPTHE 93/07

\end